\documentclass[prl,aps,twocolumn,showpacs]{revtex4}
\usepackage{graphicx}
\usepackage{amsmath}
\usepackage{amssymb}
\usepackage{bm}
\usepackage{color}

\newcommand{\be}{\begin{equation}}
\newcommand{\ee}{\end{equation}}
\newcommand{\bea}{\begin{eqnarray}}
\newcommand{\eea}{\end{eqnarray}}
\newcommand{\bei}{\begin{itemize}}
\newcommand{\eei}{\end{itemize}}

\begin{document}

\title{Adiabatic Control of Atomic Dressed States for Transport and Sensing}

\author{N. R. Cooper}
\affiliation{T.C.M. Group, Cavendish Laboratory, University of Cambridge, J.~J.~Thomson Avenue, Cambridge CB3~0HE, United Kingdom}

\author{A. M. Rey}
\affiliation{JILA, NIST \& Department of Physics, University of Colorado, 440 UCB, Boulder, CO 80309, USA}

\begin{abstract}

We describe forms of adiabatic transport that arise for dressed-state atoms in optical lattices. Focussing on the limit of weak tunnel-coupling between nearest-neighbour lattice sites, we explain how adiabatic variation of optical dressing allows control of atomic motion between lattice sites:
allowing adiabatic particle transport in a direction that depends on the internal state, and  force measurements via spectroscopic preparation and readout.
For uniformly filled bands these systems display topologically quantised particle transport.
\end{abstract}
\date{\today}


\maketitle

The topology of energy bands~\cite{hasankane,qizhang} is a concept
that has had profound influence in recent years, in the areas of both
solid state systems and ultra-cold atomic gases. In ultra-cold gases,
important experimental progress has been made in realising physics
related to the integer quantum Hall effect, by contructing
two-dimensional (2D) lattice models~\cite{goldmanreview} whose energy
bands have nontrivial topology~\cite{esslingerhaldane,munichchern}, as
characterized by a nonzero Chern number~\cite{thoulesschern}.  Indeed,
a nonzero Chern number has recently been measured in transport studies
of bosons~\cite{munichchern}.

Closely related to quantised Hall transport of 2D systems is the
quantised particle transport of (quasi)-one-dimensional (1D) systems
under time-periodic adiabatic drives. In such ``Thouless
pumps''\cite{thoulesspump}, the number of particles transported along
the 1D system is also quantised according to a Chern number, defined
over a periodic 2D parameter space spanned by the quasi-momentum
across the 1D Brillouin zone and by a time-dependent periodic
parameter varied over one cycle.

Ultra-cold gases provide an ideal setting in which to realise such
adiabatic pumping. They afford very flexible control of the lattice
potential, the possibility to vary parameters in time, and have access
to very precise probing tools~\cite{blochdz}. Although theoretical
proposals have illustrated ways to achieve quantised adiabatic
transport using optical
superlattices~\cite{wangtroyerpump,PhysRevA.90.063638,weimuellerpump}
these have been limited to far-detuned implementations that couple to
atoms in a spin-independent manner.

In this paper, we describe the new features that arise in optical
lattices involving optically dressed states of internal ``spin''
states of the atoms, within a model proposed in
Ref.~\cite{syntheticdimension} and recently realised
experimentally~\cite{sdlens,sdnist}. Although motivated by Thouless
pumping, and inheriting all features of this quantized pump, our
results will not be restricted to filled bands.  We shall emphasize a
local description which shows how adiabatic control of dressed states
can lead to novel and useful consequences. Notably, the direction of
adiabatic transport depends on the spin-state of the atom. Moreover,
the coupling of spin and orbital degrees of freedom facilitates force
measurements using only spectroscopic control. The local description
also allows one to understand in simple terms the role of inter-atomic
interactions.

\begin{figure}
\includegraphics[width=0.9\columnwidth]{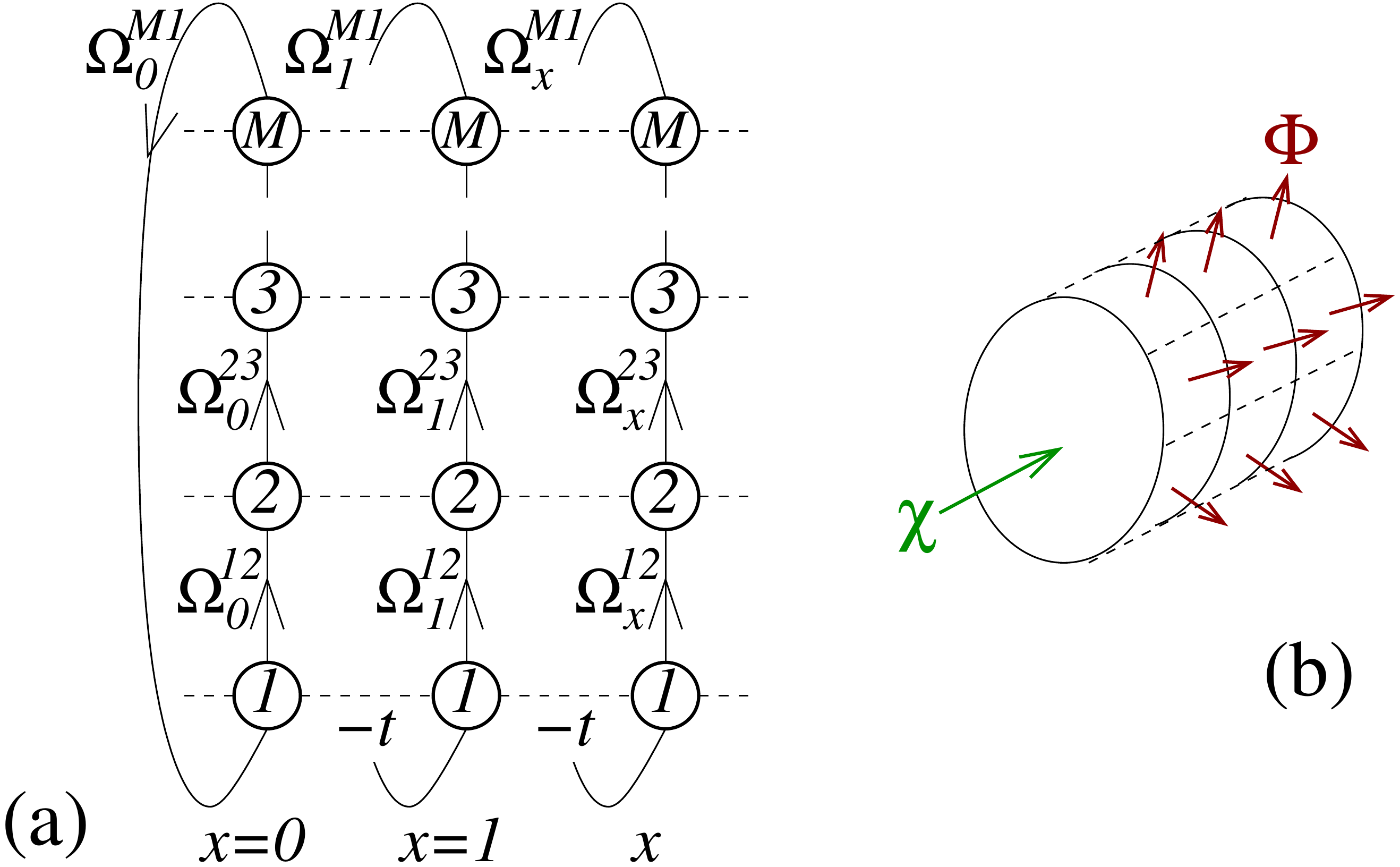}
\caption{Schematic illustration of the model. (a) The numbers denote
  the internal state $s=1,\ldots M$, spanning the vertical synthetic
  dimension. Vertical links represent the cyclic Rabi coupling.
  Horizontal links represent the tunnel coupling of neighbouring
  lattice sites, $x=\ldots, -1,0,1,2,\ldots$. (b) The coupling phases
  are such that the system can be viewed as a cylinder with a flux
  $\chi$ threading the periodic loop at $x=0$, and a flux $\Phi$
  threading each square plaquette on the surface.}
\label{fig:model}
\end{figure}

We consider a model for a spin-orbit coupled atomic gas of the form
proposed in Ref.~\cite{syntheticdimension}, which uses $M$ long-lived
internal states to implement a synthetic dimension. The model is
illustrated in Fig.~\ref{fig:model}(a).  The atoms are prepared in the
lowest band of a 1D optical lattice (for simplicity we neglect the
transverse spatial degrees of freedom, assuming this motion to be
frozen out by tight confinement). The horizontal links represent
tunnel coupling, $-t$, between neighbouring lattice sites at positions
$x=\ldots, -1,0,1,2,\ldots$, and are taken to be the same for all
internal states as is appropriate for state-independent lattices. The
vertical sites correspond to the $s=1,2,\ldots M$ internal states
which form the synthetic dimension.  As shown in
Ref.~\cite{syntheticdimension} the vertical links can be created via
hyperfine states coupled by Raman transitions in a far-detuned optical
lattice. This implementation was recently realized for $M=3$ in
Refs.~\cite{sdlens,sdnist}. Alternative implementations, allowing
larger $M$, include (magnetic sublevels of) long-lived atomic states
used in optical clocks in a ``magic" wavelength optical
lattice~\cite{Ye}.

We consider the case of  cyclic coupling where the state $s$ is coupled to both  $s-1$ and $s+1$
with $s$ interpreted modulo $M$ (i.e. $s=M+1$ is equivalent to $s=1$).  We choose the coupling from $s$ to $s+1$ to be  $\Omega_x^{s,s+1}= -\Omega e^{i\phi_{s,s+1}(x,\tau)}$ with uniform amplitude $\Omega$.
We shall require two features of the phases $\phi_{s,s+1}(x,\tau)$.
First, the phases should be spatially dependent, leading to coupling of ``spin'' and spatial degrees of freedom.
 We take
\begin{equation}
\phi_{s,s+1}(x,\tau) = \phi_{s,s+1}(0,\tau) + x \Phi\,,
\label{eq:phases}
\end{equation}for which the model maps to the Harper model in a  square lattice at ``flux'' of $\Phi$
through each plaquette. Since we consider neutral atoms it is
convenient to measure flux in dimensionless variables. Throughout we
use the convention that the ``flux'' threading any loop is the
phase picked up as a particle is transported around the loop. Hence
one flux quantum is $2\pi$.

Second, it should be possible to  vary the phase
\begin{equation}
\chi(\tau) \equiv -\sum_{s=1}^M \phi_{s,s+1}(0,\tau),
\end{equation}
in real time $\tau$ during the experiment. This phase has the simple
interpretation as the flux through the periodic loop in the synthetic
dimension at $x=0$. [See Fig.~\ref{fig:model}(b).] For $M\geq 3$
internal states $\chi$ is a non-trivial, gauge-invariant phase that
influences the spectrum.  Physically, it is set by the relative phases
of the $M$ different Rabi couplings, so is readily controllable in
experiment.  { Similar effects can arise for optical dressing of two
levels if they are resonantly coupled by a
standing wave along x (instead of running waves).
 Since the phase difference between the interfering waves displaces  the  standing wave, this case resembles in many aspects the  far-detuned
implementations~\cite{wangtroyerpump,PhysRevA.90.063638,weimuellerpump}}.

The possibility to vary the phase $\chi(\tau)$ during an experiment is
the new feature that we consider in this paper.  To make the ideas
concrete we focus on $M=4$ internal states and $\Phi = \pi/2$, but the
key features appear in more general cases. Without loss of generality,
we can choose a gauge in which the phases are uniform, with
\begin{equation}
\phi_{s,s+1}(x,\tau) = \phi(x,\tau) \equiv -\chi(\tau)/4 + x \pi/2.
\end{equation} We consider first the limit of vanishing tunnel-coupling $t=0$, for which the sites $x$ can be treated independently. The Hamiltonian describing the local Rabi couplings in the rotating wave approximation is
\begin{equation}
\hat{H}_\Omega = \sum_x \sum_{s=1}^4 \Big [\Omega_x |s+1\rangle_x\langle s|+ \Omega^*_x |s\rangle_x  \langle s+1|\Big],
\label{eq:rabi}
\end{equation} with $\Omega_x \equiv -\Omega e^{i\phi(x,\tau)}$.
The eigenstates  are the dressed states
\begin{equation}
|k_s\rangle_x = \frac{1}{2}\sum_{s=1}^4 e^{i k_s s} |s\rangle_x
\label{eq:dressedstates}
\end{equation} labelled by the allowed wavevectors along the synthetic  direction, $k_s\in \{0,\pi/2,\pi,3\pi/2\}$.
The wave functions take the same form for all $x$, but their energies vary with position according to
\begin{equation}
\epsilon_{x,k_s}  = -2\Omega \cos( k_s - x\pi/2 +\chi/4)\,.
\label{eq:siteenergies}
\end{equation} Note that the change $\chi\to \chi' = \chi + 2 m\pi$ and $k_s\to k_s' = k_s-m\pi/2$,  with $m$ an integer, leaves the spectrum unchanged and  reflects its  gauge invariance.

For isolated lattice sites, $t=0$, one can readily envisage ways to
prepare the atoms in a given dressed state. For example, this can be
accomplished by slowly ramping up the Rabi coupling $\Omega$ from zero
while keeping the lasers slightly detuned from resonance to introduce
energy offsets that are proportional to $s$. This generates the net
rotating-frame Hamiltonian $\hat{H}_\delta + \hat{H}_\Omega$, with
\begin{equation}
\hat{H}_\delta =  \delta \;\sum_x \sum_s  s |s\rangle_x  \langle s|\,.
\label{eq:detuning}
\end{equation} 
For an atom at site $x$ initially in internal state $s$, turning on
$\Omega$ slowly compared to $\delta/h$ will adiabatically transfer it
into the $s^{\rm th}$ lowest energy dressed state of $\hat{H}_\Omega$
once $\Omega \gg \delta$. (It may be advantageous to simultaneously
vary the detunings $\delta\to 0$ over this ramp.) For $\chi=0$, and
starting from $s=1$ on site $x$ this is the state with $k_s
=x\pi/2$. Reversing this protocol will allow measurements of the
dressed state occupations, since each dressed state will be
adiabatically mapped to a different internal state $s$.

Now, imagine that the atom located on site $x$ in a deep lattice,
$t=0$, has been prepared in a dressed state $|k_s\rangle_x$. Consider
reducing the lattice depth to introduce weak tunnel coupling $t\ll
\Omega$.  The tunnel coupling conserves the synthetic momentum $k_s$,
so, for typical values of $\chi$, the state $|k_s\rangle_x$ is out of
resonance from the neighbouring states, $\epsilon_{x,k_s} \neq
\epsilon_{x\pm 1,k_s}$. Since the energy offset is of order $\Omega$,
for $t\ll \Omega$ the energy eigenstates are well described by the
{\it localised} states $|k_s\rangle_x$.  We note that the periodicity
of $\epsilon_{x,k_s}$ under $x \to x+4$ requires the energy
eigenstates to be extended Bloch waves.  However, the bandwidth of
these states is of order $t^4/\Omega^3$ which for now we assume to be
small. (For $t/h \sim 100\mbox{Hz}$ and $\Omega/h = 10\mbox{kHz}$,
this bandwidth is of order $10^{-4}\mbox{Hz}\times h$.)

\begin{figure}
\includegraphics[width=0.8\columnwidth]{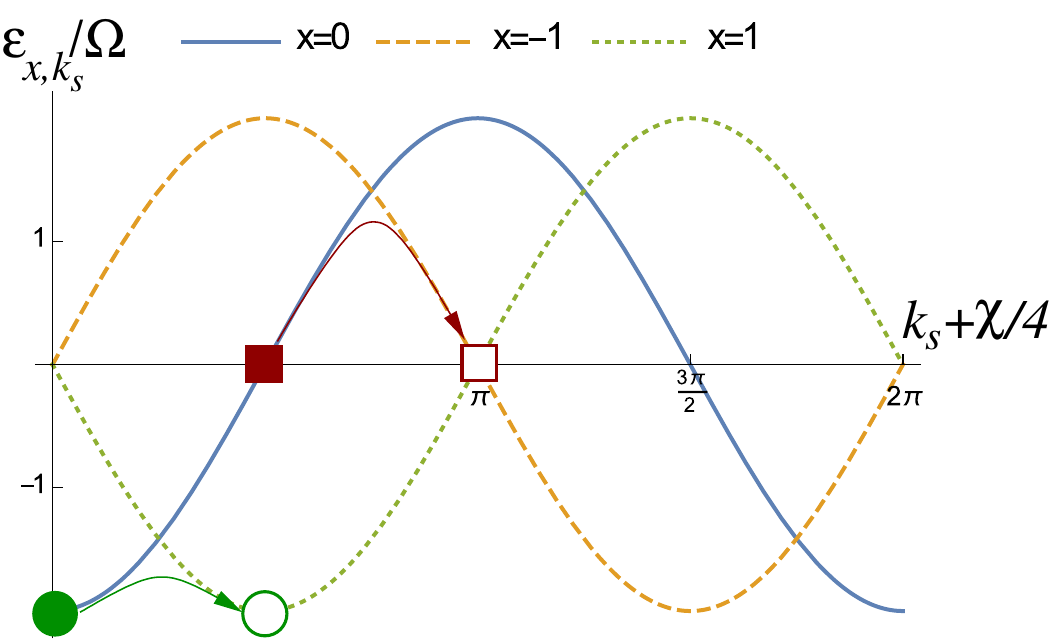}
\caption{ \label{fig:transfer} Dressed-state energies
  (\ref{eq:siteenergies}) for vanishing tunneling $t=0$, on sites
  $x=0,-1,1$ as a function of $k_s + \chi(\tau)/4$.  At the allowed
  $k_s\in \{ 0,\pi/2,\pi, 3\pi/2\}$ degeneracies between states on
  neighbouring sites, $|\Delta x| = 1$, appear only for $\chi(\tau) =
  \pi$ (modulo $2\pi$). These degeneracies are split by $t\neq 0$: a
  particle initially in state $|k_s=0\rangle_{x=0}$ at $\chi(0)=0$
  (filled circle) is then transferred adiabatically to the state
  $|k_s=0\rangle_{x=1}$ as $\chi(\tau)$ increases to $2\pi$ (open
  circle); a particle initially in $|k_s=\pi/2\rangle_{x=0}$ at
  $\chi(0)=0$ (filled square) is transferred adiabatically to
  $|k_s=\pi/2\rangle_{x=-1}$ at $\chi(\tau) = 2\pi$ (open square).  }
\end{figure}

The key feature that allows adiabatic transfer is that, by varying the
phase $\chi(\tau)$, neighbouring states can be brought into resonance
and the tunnel coupling restored.  This is illustrated in
Fig.~\ref{fig:transfer}, which shows the variation of the energy
levels (Eq. \ref{eq:siteenergies}) at sites $x=0,1$ and $-1$ as a
function of $ k_s + \chi(\tau)/4$. Consider a particle that is
prepared in the state $|k_s = 0\rangle_{x=0}$ for $\chi(0) =0$,
denoted by the filled circle in Fig.~\ref{fig:transfer}. As
$\chi(\tau)$ is increased from $0$ the energy of this state increases
smoothly until it encounters a crossing with the state
$|k_s=0\rangle_{x=1}$ at $\chi(\tau) = \pi$. For non-zero tunnel
coupling, $-t$, these two states anticross with gap $2t$.  So if
$\chi(\tau)$ is varied slowly compared to $2t/h$ the particle will
follow the ground state, ending at $\chi(\tau) =2\pi$ in the state
$|k_s=0\rangle_{x=1}$ (open circle in Fig.~\ref{fig:transfer}). Thus
the particle is adiabatically transported in the lattice, in a
direction determined by the sign of $d\chi/d\tau$. This encapsulates
the local picture of the adiabatic pumping protocol. It is a robust
process, with each particle transferred by one lattice constant as
$\chi(\tau)= \chi(0) +2\pi$, within the assumption of adiabatic
evolution.

Moreover, this adiabatic transfer has the feature that the direction
of motion depends on which dressed state the particle occupies,
$k_s$. For example, a particle starting in the state $|k_s =
\pi/2\rangle_{x=0} $ at $\chi =0$ (filled square in
Fig.~\ref{fig:transfer}) will be transferred to the state $|k_s =
\pi/2\rangle_{x=-1} $ (open square in Fig.~\ref{fig:transfer}) if
$\chi(\tau)$ is adiabatically increased to $\chi(\tau)=2\pi$. This
internal-state dependence contrasts with prior pumping protocols based
on scalar optical lattices.  It can be used as a way to separate spin
states in an adiabatic manner: while the states $k_s=0,\pi$ move to
the right, the states $k_s = \pi/2,3\pi/2$ move to the left when
$\chi(\tau) = \chi(0) +2\pi$.

These adiabatically prepared dressed states are highly sensitive to
external forces along the 1D lattice and offer the interesting
potential to detect them using {\it spectroscopy}.  Forces could arise
from external influences (e.g. gravity, or magnetic fields) or from
inter-atomic interactions.  We shall first illustrate the ideas for an
external force, $F_x$, such as gravity, that provides an
internal-state-independent energy difference $\Delta V = F_x a$
between neighbouring lattice sites ($ a$ is the lattice spacing).  A
previous proposal~\cite{carusottoforce} described a sensitive local
force sensor requiring measurements of momentum distributions, which
Bloch-oscillate at frequency $\Delta V/h$.  The dressed state approach
we present allows measurements of $\Delta V/h$ using spectroscopic
methods alone.

Note that in the above pumping protocol if the phase $\chi$ is varied
from $\chi=0$ to $\chi = \pi$ (not as far as $2\pi$), then an atom
initially in state $|k_s = 0\rangle_{x=0}$ will evolve into the state
$(1/\sqrt{2})\left[|k_s = 0\rangle_{x=0} + |k_s = 0\rangle_{x=1}
\right]$ (this in-phase combination is selected by the tunnel
coupling, $-t$).  In the presence of an additional energy offset
$\Delta V = F_x a$ between neighbouring lattice sites, adiabatic
evolution to $\chi=\pi$ loads the atom in the ground state
$|\psi\rangle_+ = \sin(\theta/2)|k_s=0\rangle_{x=0} +
\cos(\theta/2)|k_s=0\rangle_{x=1}$ where $\theta =
\sin^{-1}(t/\sqrt{(\Delta V/2)^2+t^2})$.
One can envisage various ways to extract $\Delta V$ from subsequent
measurements. One way is to measure the mean occupations
$\sin^2(\theta/2)$ and $\cos^2(\theta/2)$ of the two states
$|k_s=0\rangle_{x=0}$ and $|k_s=0\rangle_{x=1}$, which depend linearly
on $\Delta V/t$ for small $\Delta V$: $\sin^2(\theta/2) =
1-\cos^2(\theta/2) \simeq \frac{1}{2}\left[1-\Delta V/(2
  t)+\ldots\right]$.  Rapidly ramping up the 1D optical lattice to
$t=0$ freezes the particles in given lattice sites: $|k_s =
0\rangle_{x=0}$ is the local groundstate but $|k_s = 0\rangle_{x=1}$
is an excited state, so on reverting from $\chi=\pi$ to $\chi=0$ and
then removing the coupling $\Omega\to 0$ adiabatically in the presence
of the detunings (Eq. \ref{eq:detuning}) the dressed states evolve
into different internal states $s$ which are readily detected
spectroscopically. Another possibility is to start from the state
$|\psi_+\rangle$ and ramp up the lattice to suppress tunneling $t=0$
for a time $\tau_R$, during which the system performs Ramsey
oscillations between $|\psi_\pm\rangle$ at frequency $\Delta
V/h$. These can be measured once $t$ is restored by reversing the
preparation sequence. We point out however that the same sensitivity
of the system to external forces makes it vulnerable to other types of
uncontrolled noise sources (e.g. background magnetic fields) which
must be taken into account for precise metrology.

In a similar way, the adiabatic protocol can be also used to measure
inter-atomic interactions. Consider two atoms that start in the same
internal state (e.g. $s=1$) at two adjacent lattice sites
(e.g. $x=0,1$). For weak onsite interaction, $|U|\ll t , \Omega$, the
above preparation sequence and ramp to $\chi=\pi$ would place these
atoms approximately on an equal superposition of the states $|k_s=
0\rangle_{x=0}| k_s=\pi/2\rangle_{x=1}$, $| k_s=0,\pi/2\rangle_{x=1}$,
$ |k_s=0\rangle_{x=0}| k_s=\pi/2\rangle_{x=2}$, and
$|k_s=0\rangle_{x=1}| k_s=\pi/2\rangle_{x=2}$. Since there is non-zero
amplitude for both atoms to occupy $x=1$, if tunneling is suddenly
suppressed and the system is let to evolve for some time, the onsite
interactions will generate Ramsey fringes with frequency $U/h$. The
connection to force measurement with a single atom, described above,
can be made precise by filling a superlattice of double-wells, such
that only one atom is displaced at $\chi=\pi$.  Note that only SU(M)
symmetric interactions preserve $k_s$ as a good quantum number.
SU(M)-breaking interactions will further lead to detectable couplings
to states with $k_s \neq 0,\pi/2$.

We have focussed on motion and force detection in the weak tunneling
regime, $t\ll \Omega$.  For $t\sim \Omega$ the eigenstates must be
considered to be extended Bloch waves of the Harper model. They are
characterized by the 2D wavevectors $(k_x, k_s+ \chi/M)$ with
continuous $k_x$ and discrete $k_s \in\{2\pi/M \times
\mbox{integer}\}$. At flux $\Phi = (2\pi)(p/q)$, with $p$ and $q$
relatively prime integers, the Harper model has a set of energy bands
with topological character, as described by non-zero Chern number,
${\mathcal C}$~\cite{thoulesschern}.

For the 1D model considered here, ${\cal C}$ sets the number of
particles that move along the length of the system under the adiabatic
evolution of $\chi = 0 \to 2\pi$~\cite{thoulesspump}.  The resulting
quantised transport for an insulating state with an integer number,
$\alpha$, of bands filled (1D filling $n_{\rm 1D} = M \alpha/q$) is
described by the application of the iconic results of
Refs.~\cite{thoulesspump,thoulesschern}.
For $M=4$, $\Phi=\pi/2$ (i.e. $p/q=1/4$), the case $\alpha = 1$
corresponds to one particle per lattice site ($n_{\rm 1D} = 1$). The
lowest energy band of the Harper model at $t=\Omega$ has Chern number
$1$. Thus, precisely one particle transported along the 1D lattice for
each cycle $\chi(\tau) =\chi(0)+2\pi$.
This adiabatic transport is topologically protected, so is insensitive
to weak perturbations, such as interparticle interactions with
strength $|U|\lesssim t, \Omega$. For strong onsite interactions
$|U|\gg t,\Omega$, acting between all internal states $s$, the
adiabatic transport can become blocked due to the energetic
suppression of double-occupancy of any two dressed states on lattice
sites.

This topological phase of the dressed atoms can be adiabatically
prepared starting from vanishing Rabi coupling, $\Omega =0$, and a
band insulator of $n_{\rm 1D}=1$ fermion per lattice site in a single
internal state, say $s=1$. To do so, one simply ramps up the coupling
$\Omega$ of Eq.\ref{eq:rabi} in the presence of the detuning
(Eq. \ref{eq:detuning}) for $\chi\neq \pi$.  It may seem surprising
that one can adiabatically connect the trivial band insulator (at
$\Omega=0$) to an insulating state at $\Omega = t$ which is
characterised by a non-zero Chern number. However, in this 1D setting
$k_s$ is discrete, so by ramping at fixed $\chi$ the system only
explores certain lines through the 2D Brillouin zone. For $M=4$, $\Phi
= \pi/2$, for which $k_s \in \{ 0, \pm \pi/2, \pi\}$ the lowest band
only has gap closings at $k_s+\chi/4 = \pm \pi/4, \pm 3\pi/4$. For
$\chi\neq \pi$ the spectrum remains gapped and the system evolves
adiabatically.

\begin{figure}
\includegraphics[width=0.75\columnwidth]{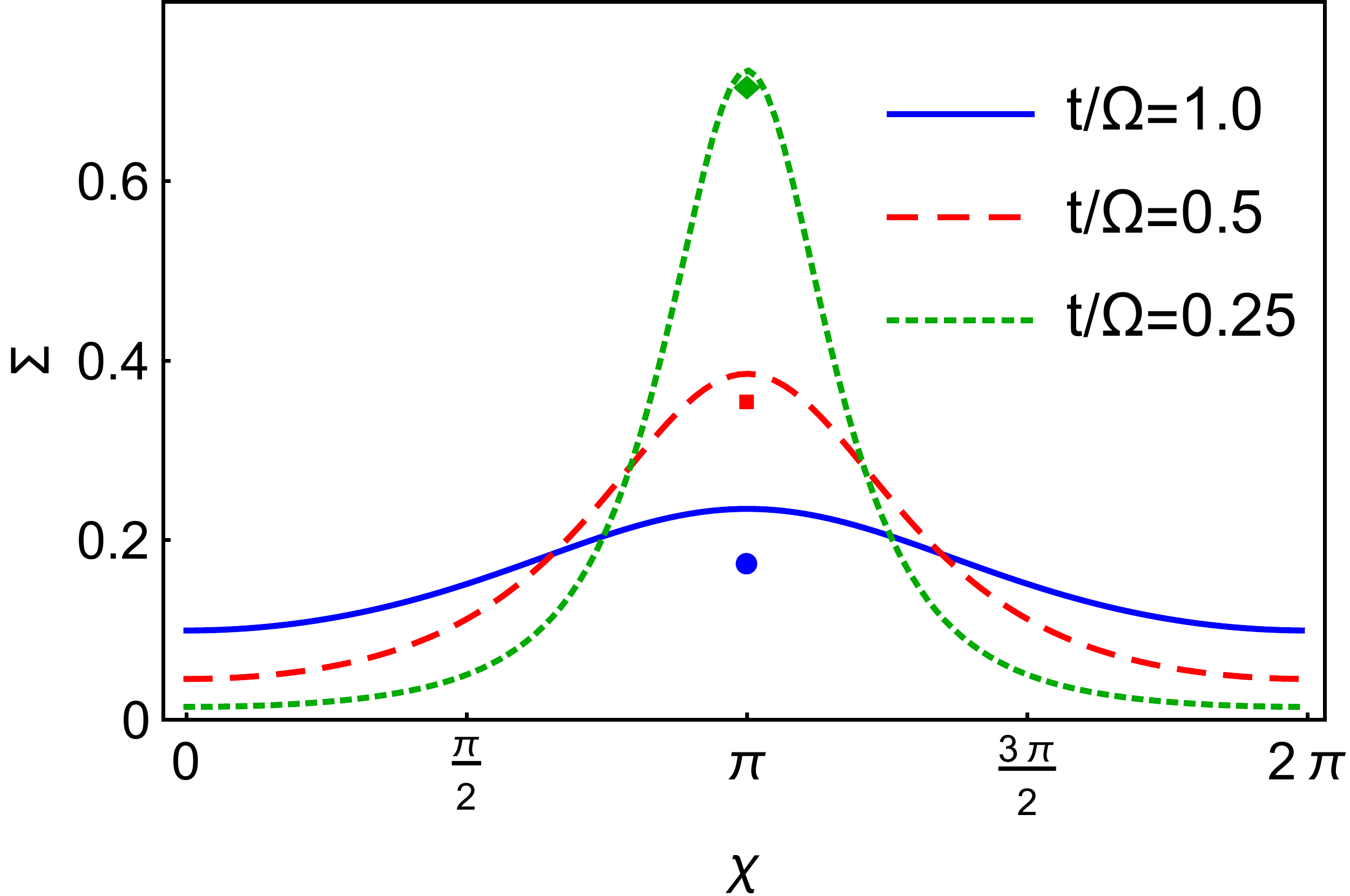}
\caption{Dimensionless conductivity $\Sigma$, describing the mean
  synthetic current $I_s$ in response to a force
  $F_x$, as a function of the phase $\chi$.
($M=4$, $\Phi = \pi/2$ and $n_{\rm 1D} = 1$ particle per lattice site.)
For weak tunneling, $t\ll \Omega$,
  the response at  $\chi = \pi$
is well-described within the local picture (points).}
\label{fig:berrycurvature}
\end{figure}

The coupling between positional motion and the dressed states allows
force detection with spectroscopic read-out also in this regime where
the energy eigenstates must be viewed within band theory~\cite{Note2}.
Consider a system of non-interacting atoms that fill a set of the
Harper bands at a fixed $\chi$ (e.g. a fermionic band insulator), or
that are uniformly distributed in $k_x$. Since the band is uniformly
occupied, a force $F_x$ does not lead to Bloch oscillations along the
$x$-direction. However, it does lead to a current along the synthetic
dimension, corresponding to a nonzero expectation value of $\hat{I}_s
\equiv \frac{1}{\hbar}\frac{\partial \hat{H}_\Omega}{\partial
  \chi}$. This arises from the existence of an anomalous velocity
associated with the Berry curvature\cite{niureview,pricecooperberry}
of the occupied states. The mean synthetic current is $I_s \equiv
\langle \hat{I}_s \rangle = - (Na/\hbar) \Sigma F_x$ with $N$ the
total number of atoms, $a$ the lattice constant, and the dimensionless
conductivity $\Sigma$ determined by the average Berry curvature along
the lines $(k_x, k_s+ \chi/M)$.  The dependence of $\Sigma$ on $\chi$
is shown in Fig.~\ref{fig:berrycurvature} for $M=4$, $\Phi = \pi/2$
and $n_{\rm 1D} = 1$.  For $t/\Omega \ll 1$ the Berry curvature is
maximum close to $\chi = \pi$, which is where bandgaps close at $t\to
0$.
While at any given $\chi$ this conductivity is not quantized, its
integral $\int_0^{2\pi} \Sigma \;d\chi$ is the (integer) Chern number.
Note that the eigenstates of $\hat{I}_s$ are the same as those of
$\hat{H}_\Omega$ (Eq. \ref{eq:rabi}), given by
Eq. \ref{eq:dressedstates}.  Thus, their occupations --- and therefore
the mean synthetic current when weighted by the eigenvalues $I_{x,k_s}
= (\Omega/2\hbar)\sin(k_s-x\pi/2 +\chi/4)$ --- can be measured by the
adiabatic ramps described before, in which dressed states
adiabatically return to different internal states $s$.
For weak tunneling $t\ll \Omega$ this reduces to the two-state problem
described above in the local description. The linear dependence of
$\sin^2(\theta/2)$ on $\Delta V = F_x a$ for $\chi =\pi$ corresponds
to a Berry-curvature induced synthetic current, $I_s = \frac{\Omega}{2
  \hbar}\frac{1}{\sqrt{2}} \sum_x \left[\sin^2 (\theta/2) - \cos^2
  (\theta/2)\right] = - \frac{\Omega}{2\sqrt{2}\hbar} \frac{\Delta
  V}{2t} \times N = - \frac{Na}{\hbar} \frac{\Omega}{ 4 \sqrt{2}t}
F_x$ with $N$ the number of atoms.  This limiting result, $\Sigma =
\frac{\Omega}{4\sqrt{2}t}$, is shown as points in
Fig.~\ref{fig:berrycurvature}, accurately describing $\Sigma$ for
$t/\Omega\ll 1$ \cite{Note3}.

In summary we have described protocols for the spectroscopic control
of atomic dressed states that allow for adiabatic transport with
internal-state dependence, and force detection with spectroscopic
readout.  These features arise even on a {\it local} level, not
requiring extended Bloch waves. For filled bands the set-up provides a
direct realisation of a Thouless pump.  Although we assumed
translationally-invariant systems, the protocols are robust to
including a trapping potential which can further facilitate the
observation of particle pumping by the introduction of atomic cloud
edges where particles can accumulate.

\vskip0.1cm

\acknowledgments{The authors thank Jun Ye and Michael Wall for useful
  discussions. This work was supported by EPSRC Grant EP/K030094/1, by
  the JILA Visiting Fellows Program, the NSF (PIF-1211914 and
  PFC-1125844), AFOSR, AFOSR-MURI, NIST and ARO individual
  investigator awards.}

\bibliographystyle{prsty}

\end{document}